\documentclass[reprint,
superscriptaddress,
amsmath,amssymb,
aps,
floatfix,
]{revtex4-2}

\DeclareUnicodeCharacter{2009}{\,} 
\usepackage{graphicx}
\usepackage{bm}
\usepackage{hyperref}
\hypersetup{colorlinks=true, urlcolor=blue, linkcolor=blue, citecolor=blue}
\usepackage{amssymb}
\usepackage{amsthm}
\usepackage{amsmath}
\usepackage{xcolor}
\usepackage{comment}
\usepackage{gensymb}

\DeclareMathOperator\erf{erf}
\begin{document}

\title{Pressure-driven homogenization of lithium disilicate glasses}%

\author{Yasser Bakhouch}
\affiliation{LS2ME, Facult\'{e} Polydisciplinaire Khouribga, Sultan Moulay Slimane University of Beni Mellal, B.P 145, 25000 Khouribga, Morocco}
\author{Silvio Buchner}%
\email{silvio.buchner@ufrgs.br}
\author{Rafael Abel Silveira}
\author{Leonardo Resende}
\author{Altair Soria Pereira}
\affiliation{Graduate Program in Materials Science, Federal University of Rio Grande do Sul, Porto Alegre, Brazil}
\author{Abdellatif Hasnaoui}
\affiliation{LS2ME, Facult\'{e} Polydisciplinaire Khouribga, Sultan Moulay Slimane University of Beni Mellal, B.P 145, 25000 Khouribga, Morocco}
\author{Achraf Atila}
\email{achraf.atila@uni-saarland.de}
\affiliation{Department of Material Science and Engineering, Saarland University, Saarbr\"{u}cken, 66123, Germany}


\begin{abstract}
Lithium disilicate glasses and glass-ceramics are good potential candidates for biomedical applications and solid-state batteries, and serve as models of nucleation and crystal growth. Moreover, these glasses exhibit a phase separation that influences their nucleation and crystallization behavior. The atomistic mechanisms of the phase separation and their pressure dependence are unclear so far. Here, we used molecular dynamics simulations supported by experiments to assess the spatial heterogeneity of lithium disilicate glasses prepared under pressure. We show that the glass heterogeneity decreases with increasing the cooling pressure and almost disappears at pressures around 30 GPa. The origin of the heterogeneity is due to the attraction between Li cations to form clustering channels, which decreases with pressure. Through our results, we hope to provide valuable insights and guidance for making glass-ceramics with controlled crystallization. 
\end{abstract}

\maketitle


\section{\label{Introduction}Introduction}
Oxide glasses are pivotal materials with a wide unique combination of properties that are pertinent for functional~\cite{Chester2022}, structural~\cite{Wondraczek2011}, and technologically relevant applications~\cite{Wondraczek2022}. Silicate-based glasses, as an example, are used as ultrathin substrates, strong transparent cover materials, or ion-conducting glasses~\cite{Wondraczek2022, Wondraczek2011}.
The structure of silica glass is known to have a three-dimensional, fully connected structure made of SiO$_4$ tetrahedra~\cite{Ganisetti2023}. With the addition of modifiers such as Li$_2$O, the Si--O--Si bridging bonds break, and the bridging oxygen (BO) atoms are transformed to non-bridging oxygen (NBO) atoms, leading to depolymerization of the structure and a decrease of the glass transition temperature (T\textsubscript{g}) ~\cite{Atila2019a, Atila2020a}. The degree of depolymerization of the structure depends on the type and content of the modifier, and affects the glass properties, such as mechanical and optical properties, as well as its crystallization behavior~\cite{Atila2019b, Ouldhnini2021, Kuwik2020}.
The control of the crystallization enabled the formation of glass-ceramics, which are materials that combine the properties of both glasses and ceramics~\cite{Deng2021, Serbena2015, Fu2017, Lschmann2022}. These glass-ceramics have a unique combination of high strength and toughness, bioactivity, as well as excellent thermal stability. This set of properties made them useful in dental and medical applications, such as dental crowns and implants~\cite{MOBILIO2017, ROJAS2022}.

The advent of glass-ceramics dates back to the 1950s in the pioneering work of Stookey~\cite{Stookey1959}, wherein lithium disilicate (LS$_2$) served as the basis for the first glass-ceramics compositions. This discovery of glass-ceramics led to many open questions pertaining to the mechanisms underlying the nucleation and crystallization, as well as strategies for enhanced control over these processes while considering sample composition and thermal treatment~\cite{Allix2019}.

Several authors investigated the effect of processing on the properties of lithium silicate glasses and glass-ceramics. For instance, Kitamura \textit{et al.}~\cite{KITAMURA2000} studied lithium disilicate glass densification under high pressure. They found that subjecting the glass to high pressures resulted in an increase in the density of the glass. This densification was conjectured to be related to the distortion and increase in the packing density of SiO$_4$ tetrahedra. 
Voigt \textit{et al.}~\cite{Ulrike05} investigated the local structure and spatial distribution of lithium ions in lithium silicate glasses using nuclear magnetic resonance (NMR) and molecular dynamics (MD) simulations. Their findings indicated significant cation clustering, with stronger dipolar fields observed at the tetrahedra with three BOs, otherwise known as Q$^3$ sites in the Q$^n$ notation where $n$ denote the number of BO per tetrahedra, compared to Q$^4$ sites. Molecular dynamics simulations confirmed these conclusions, particularly in glasses with low lithia contents (x = 0.10 and 0.17), which tend to show a phase separation. 
Habasaki \textit{et al.}~\cite{Habasaki13}, used molecular dynamics simulations to investigate the structure of lithium disilicate glasses and melts under different pressures. Their results showed that the Q$^n$ units changed during the compression or decompression process, with the highest percentage of Q$^3$ structures occurring near atmospheric pressure around T$_g$. These changes in Q$^n$ distribution were driven by variations in volume or pressure and were attributed to differences in the volumes of structural units. 
Recently Bradtm\"{u}ller \textit{et al.}~\cite{Bradtmller2022}, in a combined NMR and MD study, reported atomic-scale structural insights into the effect of sub-T$_g$ annealing on the structural relaxation of LS$_2$ glasses. They confirmed that the arrangement of the network modifier in homogeneously nucleating glasses is closely related to those in the isochemical crystals, which was hypothesized by Zanotto \textit{et al.}~\cite{Zanotto2015}.

On the other hand, Buchner \textit{et al.}~\cite{BUCHNER2011, BUCHNER2013, BUCHNER2014, BUCHNER2021}, in a series of studies, investigated the effect of processing conditions on the crystallization and properties of lithium silicate glasses using different experimental techniques. They found that processing lithium silicate glasses at high temperature and pressure increased Young's modulus and hardness by around 45\% and 25\%, respectively, compared to the pristine glass. Moreover, they showed that the lithium metasilicate (Li$_2$SiO$_3$) phase can form from lithium disilicate glasses if processed under high pressure and temperature. The Li$_2$Si$_2$O$_5$ phase was stable up to 6 GPa while the Li$_2$SiO$_3$ phase began to form at 4.25 GPa and coexisted with quartz and coesite up to 6.5 GPa. The lack of lithium in the initial lithium disilicate composition led to the segregation of SiO$_2$ phases during the crystallization of Li$_2$SiO$_3$~\cite{BUCHNER2013}. 
Another important result found in Buchner \textit{et al.}~\cite{BUCHNER2014} work is the formation of a distinctive amorphous phase of densified LS$_2$ glass at a pressure of 7.7 GPa, wherein small distortions on the local arrangement of the initial structure cannot be used to explain its appearance. Moreover, Resende \textit{et al.}~\cite{Resende2021} showed that it is possible to produce lithium disilicate glasses at high pressure by melting and quenching them in situ at a pressure of 7.7 GPa. The results showed a significant increase in mechanical properties and completely different thermal properties compared to a reference glass sample conventionally prepared at atmospheric pressure. 

Although experiments provided great details on the effect of processing on the structural and physical properties of lithium disilicate glasses, the atomic-scale picture of the events that govern the change in the properties is unclear. Molecular dynamics (MD) simulations provide quantitative and qualitative insights at the atomic scale that can be used to understand these mechanisms better. In this context, we study the effect of pressure on the densification mechanisms of lithium disilicate glasses, and we provide a detailed analysis of the glass structure at both short- and medium-range. The structural properties of the glass are compared with the experimental data of glasses similar to those prepared by Resende \textit{et al.}~\cite{Resende2021} and analyzed through synchrotron X-ray diffraction (XRD).

The remainder of this paper is organized as follows: In Sec.~\ref{Sec:Method}, we describe the procedure followed to obtain the results. The calculated properties are presented in Sec.~\ref{Sec:Results}. In Sec.~\ref{Sec:Discussion}, we discuss the results and suggest possible explanations for the obtained results. Concluding remarks are given in Sec.~\ref{Sec:Conclusion}.

\section{\label{Sec:Method}Methodology}
\subsection{Experimental procedure}
Four lithium disilicate glass samples were prepared using the procedure described below. One glass batch was produced at atmospheric pressure, and the other three samples were processed under high pressure in different conditions. Standard reagent grade Li$_2$CO$_3$ (Aldrich Chem. Co. $\geq$ 99\%) and silicon dioxide (SiO$_2$), (Sigma-Aldrich, $\geq$~99.9\%) were used to produce all samples. For the glass batch produced at atmospheric pressure, SiO$_2$ and Li$_2$CO$_3$ reagents were dried in a furnace for 2 h at 120 \degree C, weighed, mixed, and melted in a Pt crucible at 1500 \degree C for 2 h in an electric furnace. The melt was poured on a steel plate at room temperature. Afterward, the batch was annealed at 430 \degree C for 1 h and left to cool slowly to room temperature. This glass is called “pristine glass” in the following. Toroidal-type high-pressure apparatus was used to process samples at high pressure and high temperature (HPHT), according to the methodology described in~\cite{sherman1987experimental, Khvostantsev2004, eremets1996high}. The pressure was initially increased to 7.7 GPa, and then the temperature was raised to 1873 K to melt the sample. Then the sample was maintained at these conditions of high temperature and pressure for 10 min and subsequently quenched within 3 min while maintaining the pressure. After this procedure was done, the pressure was slowly released.

Finally, after HPHT processing, the surface of the samples was grounded on SiC abrasive paper up to \#1500 and polished with CeO$_2$ slurry for all analyses. 

\subsection{X-ray diffraction}
Transmission geometry synchrotron X-ray diffraction (XRD) measurements were carried out at 20 keV ($\lambda$ = 0.6199~\text{\AA}) on the XDS beamline of the Brazilian Synchrotron Light Laboratory (LNLS). The powder samples were enclosed within 0.63 mm Kapton capillaries and rotated throughout the data collection.
The total structure factor S(K) for an amorphous phase containing N chemical elements is derived from the normalized scattered intensity on a per-atom scale~\cite{Faber1965}, denoted as I$_a$(K), following the equations:
\begin{equation}
  S(K) = \frac{I_a - [<f^2(K)>-<f(K)>^2]}{<f(K)>^2}
\end{equation}

\begin{equation}
S(K) = \sum_{i=1}^N \sum_{j=1}^N W_{ij}S_{ij}(K)
\end{equation}

where k = 4$\pi/\lambda$ sin($\theta$) is the transferred momentum and $W_{ij}(K)$ represents the weights of the partial structure factors and can be obtained from:
\begin{equation}
W_{ij}(K) = W_{ji}(K) = \frac{c_ic_jf_i(K)f_j(K)}{<f(K)>^2}
\end{equation}
where c$_i$ is the concentration of $i$ atoms and $f$ is the atomic scattering factor and can be calculated from the prime scattering factors~\cite{Faber1965}.
From the structure factor, the total pair distribution function, $G(r)$, can be calculated from a Fourier transform taking into account the atomic density, $\rho_0$, using the equation:
\begin{equation}
G(r) = 1 + \frac{1}{2\pi \rho_0 r}\int_0^\infty K[S(K) - 1]\sin(Kr)dK
\end{equation}
\subsection{Simulation Details}
LAMMPS code~\cite{LAMMPS} was used to perform the classic molecular dynamics simulations of LS$_2$ glasses. The equation of motion was integrated using the velocity Verlet algorithm as implemented in LAMMPS, with a time step of 1 fs. The interactions between atoms part were modeled by the Pedone \textit{et al.}~potential~\cite{Pedone2006}. The potential parameters and partial charges are provided in Ref.~\cite{Pedone2006}. An interaction cutoff of 5.5 \text{\AA} was used for short-range interactions, while the long-range interactions were treated by adopting Fennell damped shifted force (DSF) model~\cite{Fennell2006}, with a damping parameter of 0.25 \text{\AA}$^{-1}$ and of 8.0 \text{\AA} as a long-range cutoff. This potential gives a realistic agreement with available experimental data as mentioned in the literature~\cite{Atila2020a, Atila2019b, Ouldhnini2021, Atila2022, Ouldhnini2022, Atila2020c, Ghardi2019, Luo2021}, as it was designed to reproduce the structural and mechanical properties of a wide range of oxide glasses. 

\begin{figure}[h!]
\centering
\includegraphics[width=\columnwidth]{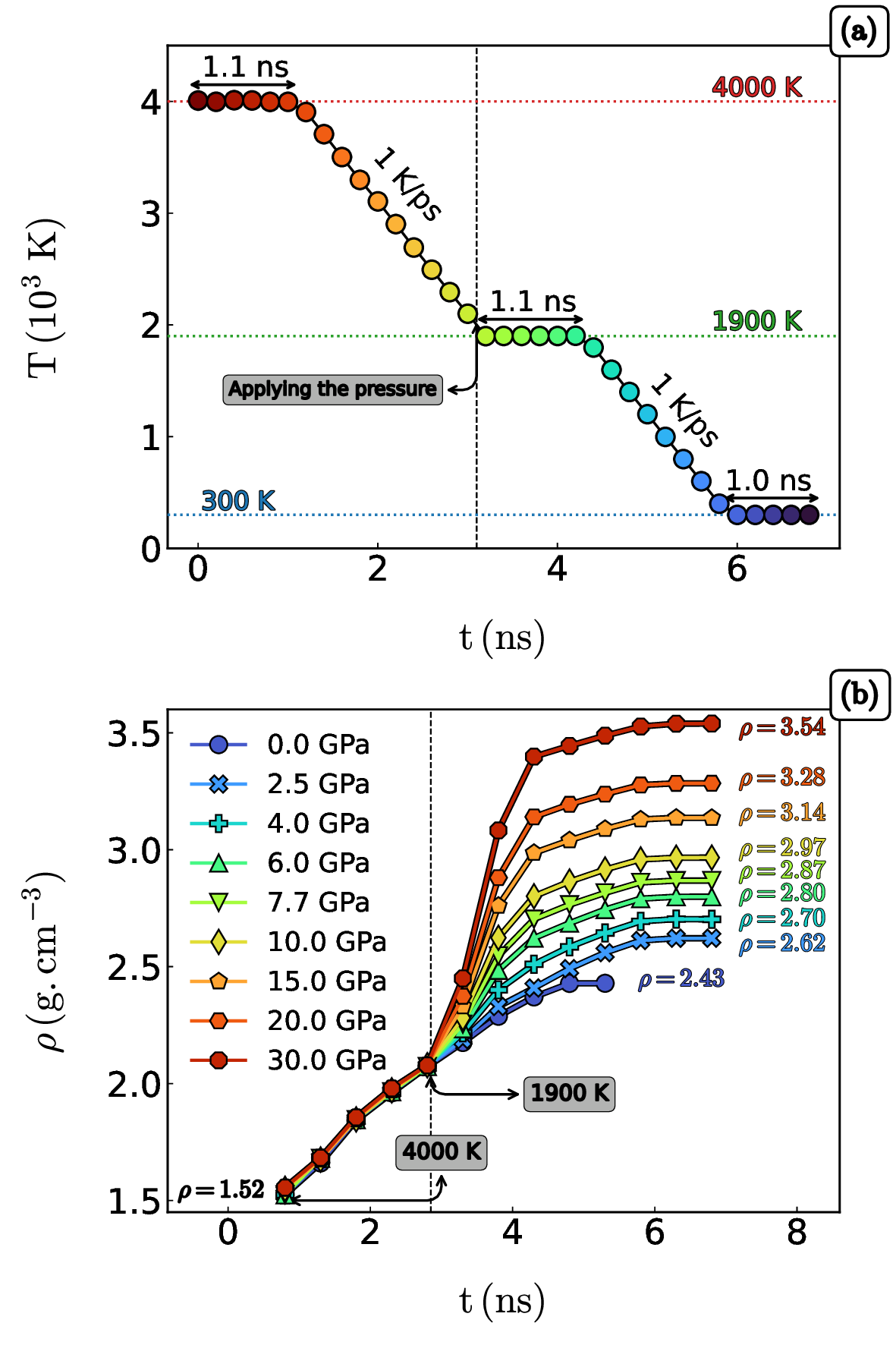}
\caption{(a) the simulation profile used to obtain the glasses, (b) the evolution of density as a function of simulation time for the different prepared glasses.}
\label{fig:den}
\end{figure}

Several LS$_2$ glasses of stoichiometric composition (Li$_2$O)$_{0.33}$--(SiO)$_{0.67}$ samples were produced by randomly placing 30000 atoms in a 3-D periodic cubic simulation box while ensuring that there are no overlapping atoms. The steps followed to cool the glasses are schematically illustrated in Fig.~\ref{fig:den}(a) and will be briefly summarized in the following. The samples were equilibrated at a high temperature (4000 K) for 100 ps in the NVT ensemble and for 1 ns in the NPT ensemble, which was enough to make the samples lose the memory of the initial configuration. The melt was then cooled from 4000 K to 1900 K in NPT with zero external pressure. At 1900~K a pressure (P) was applied (i.e., P = 2.5, 4, 6, 7.7, 10, 15, 20, and 30 GPa) for 1 ns. The samples were held at the same pressure for 100 ps and then quenched to room temperature while maintaining the pressure conditions. All cooling simulations were performed using a cooling rate of 1 K/ps. After quenching, the glass was further equilibrated at 300 K in the NPT ensemble for 1 ns. Finally, we obtained seven LS$_2$ glasses with different densities that depend on the pressure: 2.43, 2.62, 2.70, 2.80, 2.87, 2.97, 3.14, and 3.54 g/cm$^{3}$ (see Fig.~\ref{fig:den}(b)). 

\section{\label{Sec:Results}Results}
\subsection{X-ray diffraction structure factor}
\begin{figure}[!ht]
\centering
\includegraphics[width=0.98\columnwidth]{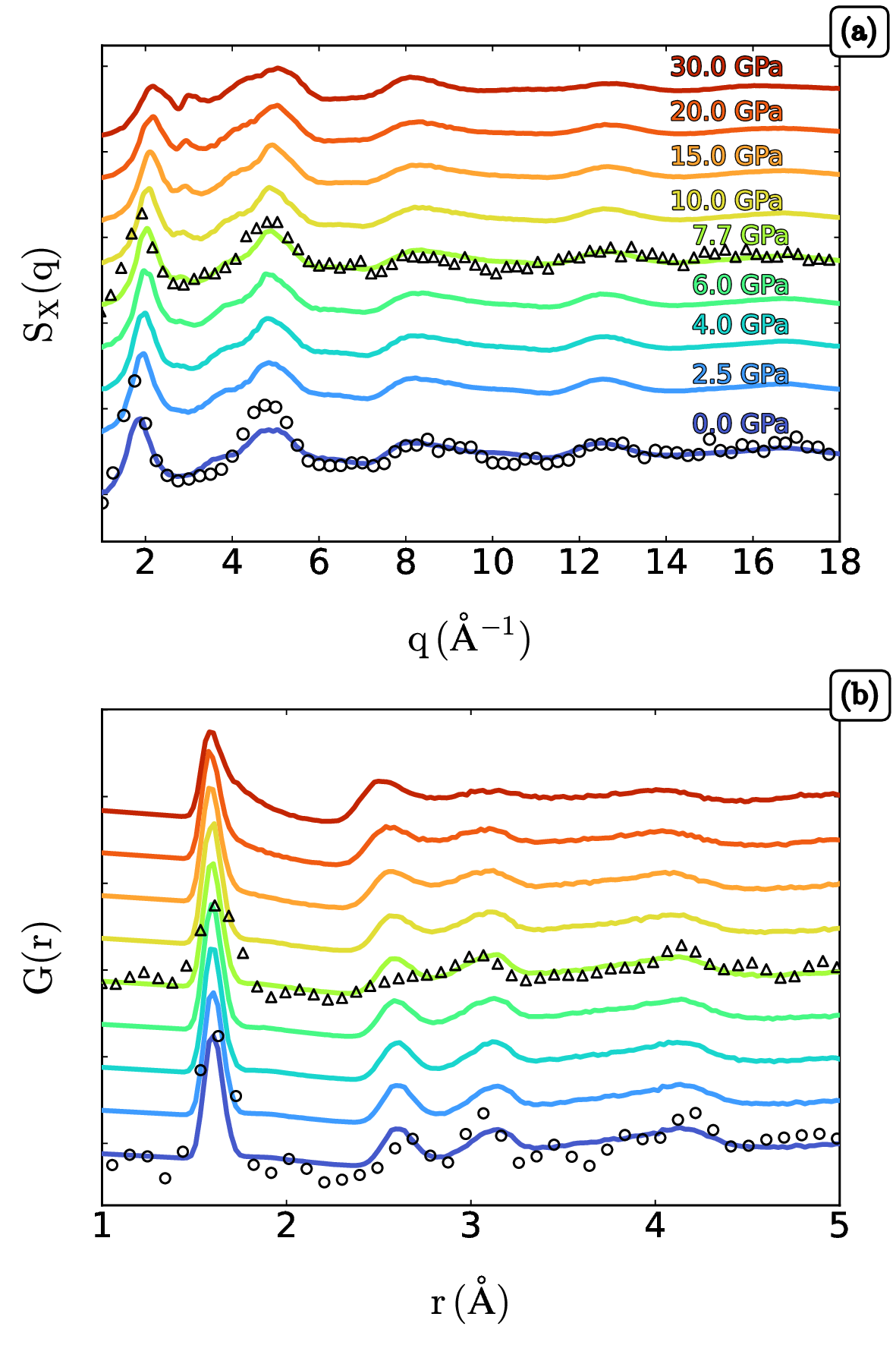}
\caption{(a) X-ray diffraction structure factors for the pristine glass and all HPHT processed samples from both experiments (open symbols) and MD simulations (solid lines); (b) calculated G(r) from the structure factor data.}
\label{fig:xrsf}
\end{figure}

Figure.~\ref{fig:xrsf}(a) shows the X-ray structure factor obtained from the simulated lithium disilicate glasses at different pressures. The structure factors obtained through X-ray diffraction experiments for the pristine glass and the one prepared at a pressure of 7.7 GPa are shown in open symbols in the same figure for comparison. The peak position and the overall shape of the structure factor are well reproduced in samples prepared by MD simulations.

From the X-ray structure factor (S$_X$(K)), we obtained the total pair distribution function, G(r), which was calculated and can be seen in Fig.~\ref{fig:xrsf}(b), the data from the MD simulations at different pressure is also shown in the same figure. Similarly to the structure factor, the G(r) of the simulated glasses compares well with that obtained by X-ray diffraction. It is important to remember that the simulation results were obtained under pressure, while the experimental ones were obtained at atmospheric pressure for a sample that was processed at 7.7 GPa, which causes slight differences in the peak positions. 

\subsection{Neutron diffraction structure factor} 
\begin{figure}[!ht]
\centering
\includegraphics[width=0.98\columnwidth]{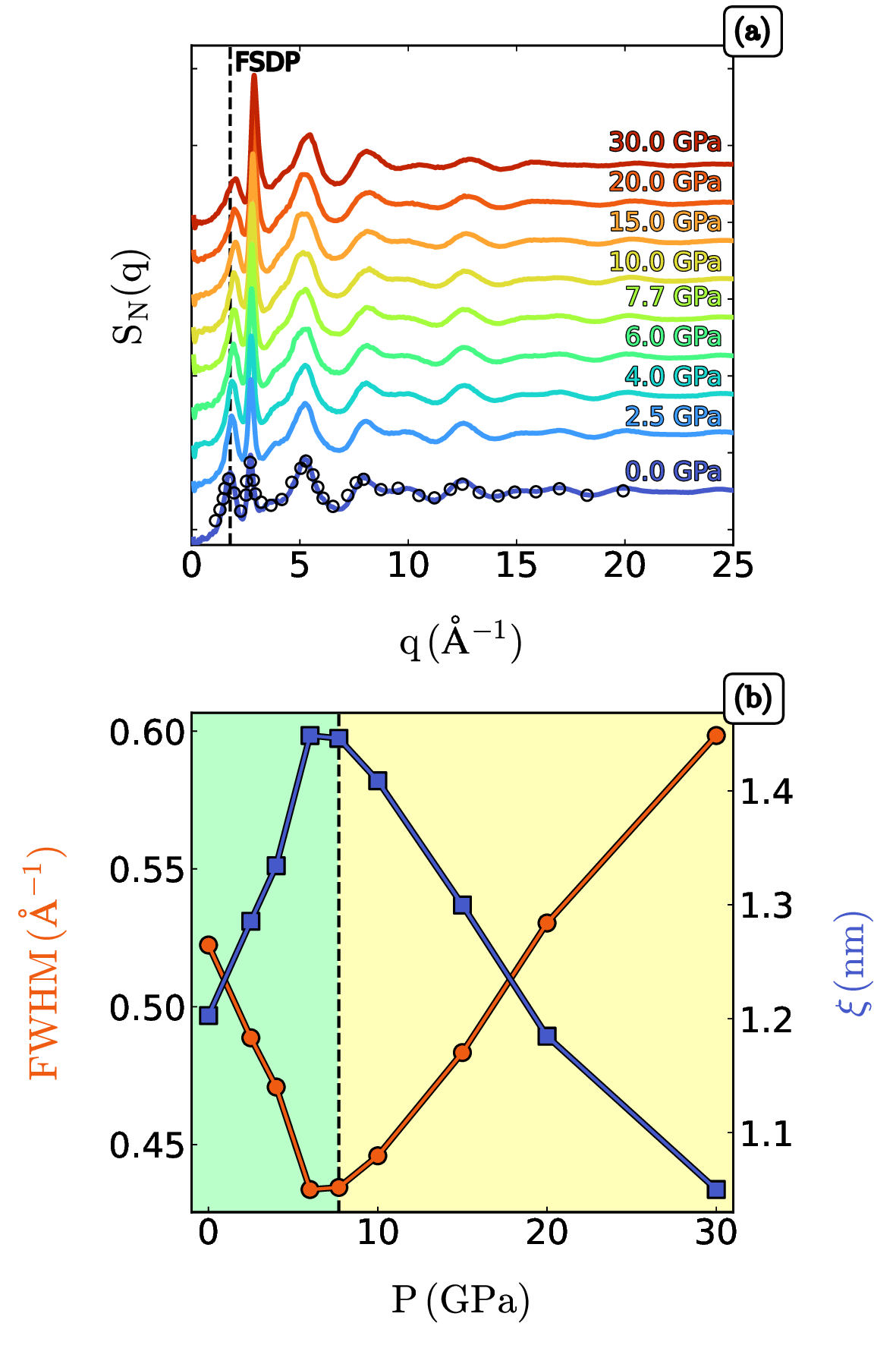}
\caption{(a) Total structure factor as a function of pressure at 300 K. (b) The full width at half maximum of the first-sharp diffraction peak of the total structure factor as a function of the pressure (left y-axis) and the corresponding correlation length (right y-axis) as a function of the pressure. Error bars are smaller than the symbol size}
\label{fig:sf}
\end{figure}

The neutron structure factor of the glass prepared under zero external pressure is compared to the one measured experimentally by neutron scattering experiments from the work of Kitamura \textit{et al.}\cite{KITAMURA2000}, and shown in Fig.~\ref{fig:sf}(a). 
The structure factor of the simulated LS$_2$ glass at zero pressure compares well to that obtained experimentally, where the positions of the first, second, and third peaks are well reproduced at 1.78, 2.72, and 5.24 $\text{\AA}^{-1}$. 
This good agreement highlights the interatomic potential's ability to simulate glass models with realistic glass structures. Moreover, similarities are observed in the intensity of the peaks, especially that of the first sharp diffraction peak (FSDP) in both MD and experiments. The overall shape of $S(q)$ obtained from experiments is well reproduced in the MD simulated glass structures, with slight deviations (at large $q$ values) that can arise from the high cooling rate used in MD simulations.
The pressure effect was mainly observed in the low-$q$ region of the structure factor (See Fig.~\ref{fig:sf}(a)), which suggests that the pressure does not significantly affect the short-range structure (at least at small pressures). 
The shift towards higher $q$ values of almost all peaks with increasing pressure highlights the compaction of the network as the pressure increases. It can be seen that the FSDP shift towards high $q$ values with increasing pressure, and its
intensity decreases with increasing pressure. It is worth mentioning that this behavior of the FSDP is in good agreement with experimental observations and also results of simulated close glass compositions~\cite{KITAMURA2000, Uhlig96, DU06,
LE20}. 
The FSDP, which corresponds to structural correlations on a larger length scale, was fitted using a skewed Gaussian function, and the full width at half maximum (FWHM) of the FSDP was obtained. From this FWHM, we can extract a correlation length in the medium range, allowing us to understand the correlations in the medium-range order. 
The correlation length was calculated using eq.~\ref{eq:FWHM}~\cite{Atila2020b, Bauchy17, Du05, Elliott91}.
\begin{equation}
\label{eq:FWHM}
\xi = \frac{2\pi}{FWHM}
\end{equation}
Figure.~\ref{fig:sf}(b) shows the variation of FWHM and the correlation length as a function of pressure. The FWHM increases with increasing pressure up to pressures around 6.0 GPa, then decreases to values lower than those obtained for the glass prepared under no external pressure, highlighting the possibility of polyamorphism. The correlation length behaves inversely to the FWHM. When the FWHM increases, the correlation length decreases, and vice versa. The increase in the correlation length indicates an increase in a correlated structure at the intermediate range.

\subsection{Short-range structure}
\begin{figure}[!ht]
\centering
\includegraphics[width=0.98\columnwidth]{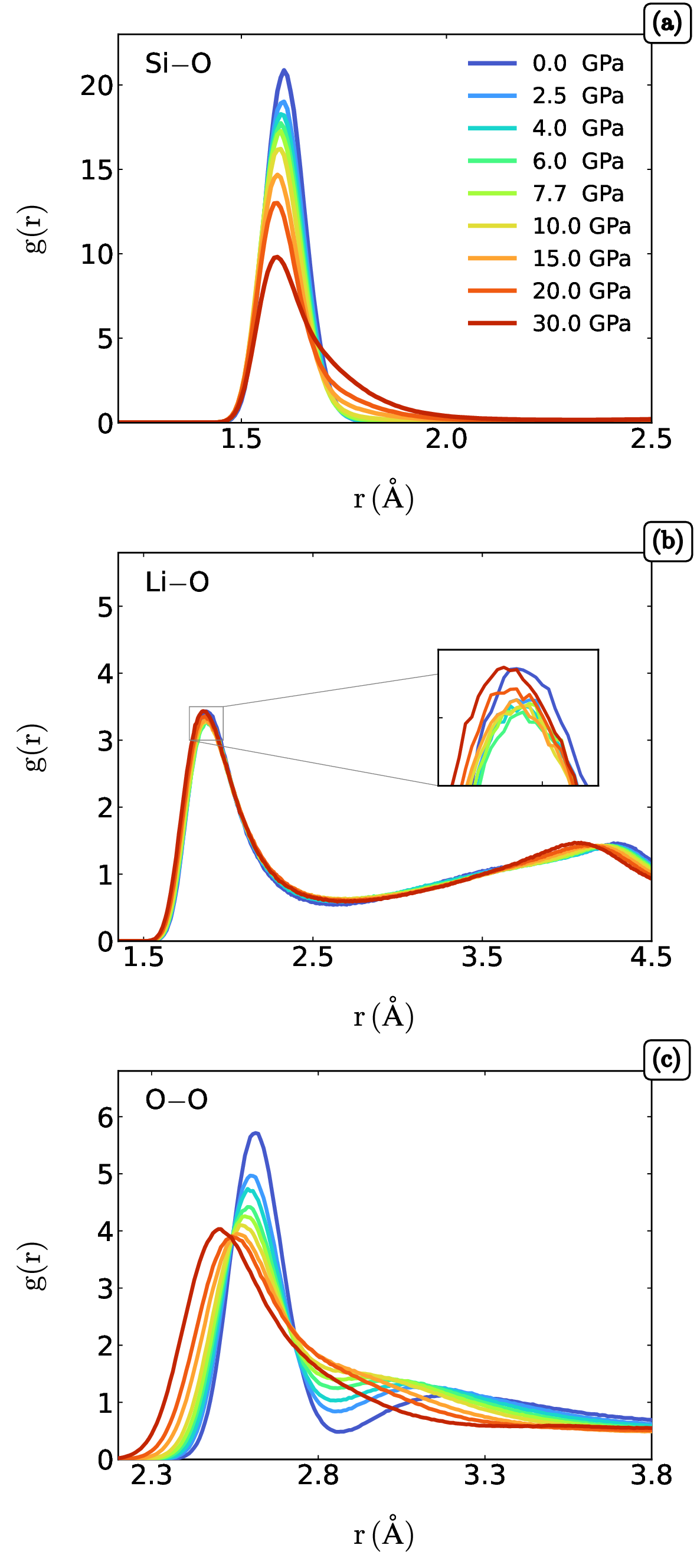}
\caption{(a) Si--O, (b) Li--O, and (c) O--O partial distribution functions (g(r)) as a function of pressure in LS$_2$ glasses at 300 K.}
\label{fig:PDF_CN}
\end{figure}

The short-ranged structure of the glass is defined up to the first coordination shell of the cations with the oxygen atoms. The Partial distribution function (g(r)) describes the short-range structure, while the location of the first peak in the g(r) for the corresponding elements determines their nearest neighbor distance. Figure.~\ref{fig:PDF_CN} illustrates the g(r) of Si--O, Li--O, and O--O plotted up to their first minima. 
The first peak of Si--O pair shifts towards shorter distances, its intensity decreases, and becomes broader with increasing pressure. The Li--O g(r) first peak profile showed no noticeable change with increasing pressure. The O--O g(r) behaves similarly to the Si--O g(r), with peaks shifting to lower $r$ values and the first peak becoming broader. 

\begin{figure}[!ht]
\includegraphics[width=\columnwidth]{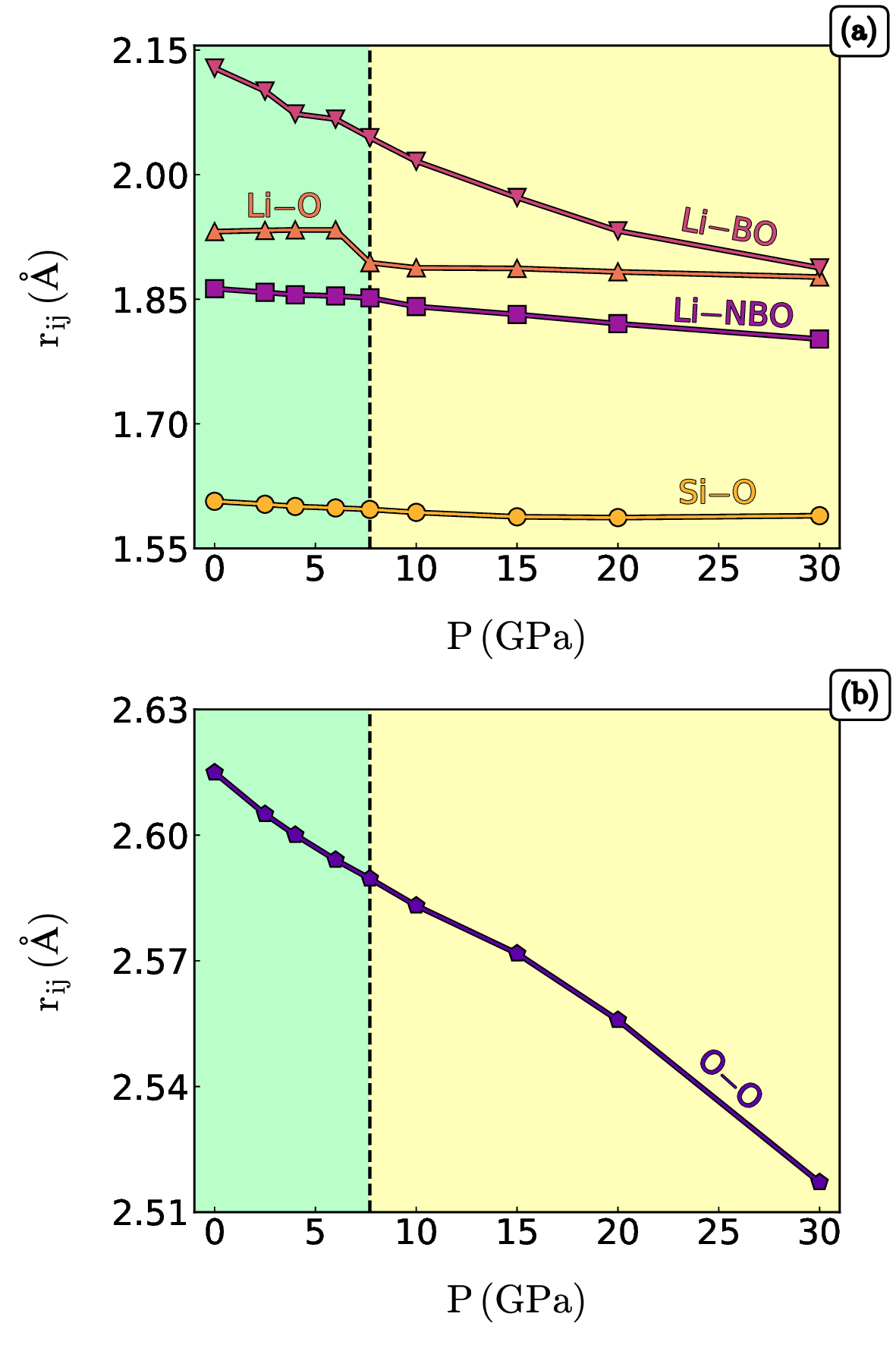}
\caption{(a) Si--O and Li--O, (b) O--O bond lengths as a function of pressure in LS$_2$ glasses at 300 K. Error bars are smaller than the symbol size.}
\label{fig:BL}
\end{figure}

\begin{figure}[!ht]
\includegraphics[width=.98\columnwidth]{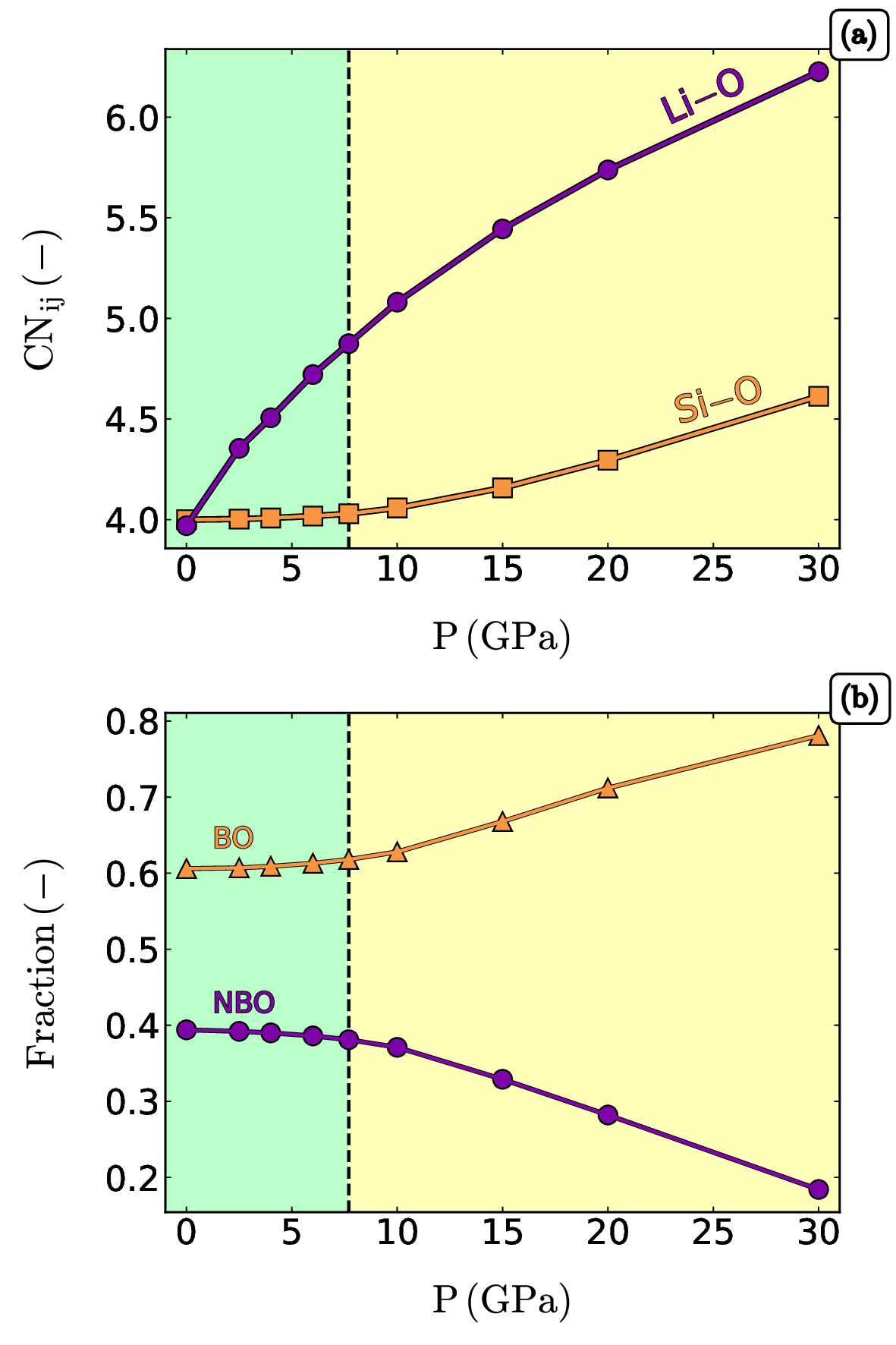}
\caption{(a) Si--O and Li--O coordination numbers, (b) O--O coordination numbers, and (c) oxygen species. as a function of pressure in LS$_2$ glasses at 300 K. Error bars are smaller than the symbol size.}
\label{fig:CN}
\end{figure}

\begin{figure}[ht]
\centering
\includegraphics[width=\columnwidth]{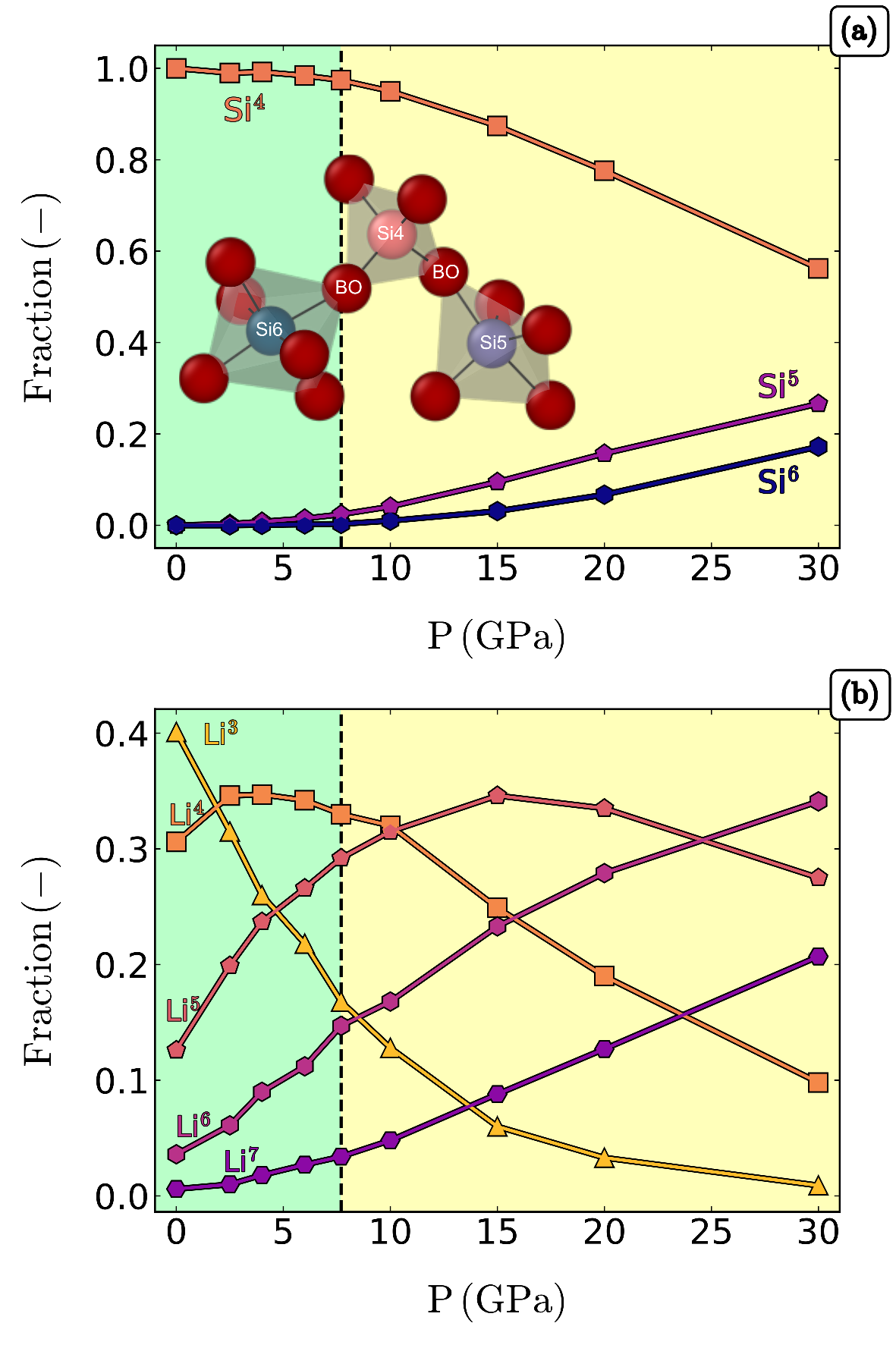}
\caption{(a) Si$^n$ coordination state, with a snapshot of 4, 5, and 6-fold coordinated Si atoms. (b) Li$^n$ coordination state (n denotes the number of O neighbors atoms), as a function of pressure in LS$_2$ glasses at 300 K. Error bars are smaller than the symbol size}
\label{fig:xion}
\end{figure}

\begin{figure}[h!]
\includegraphics[width=.98\columnwidth]{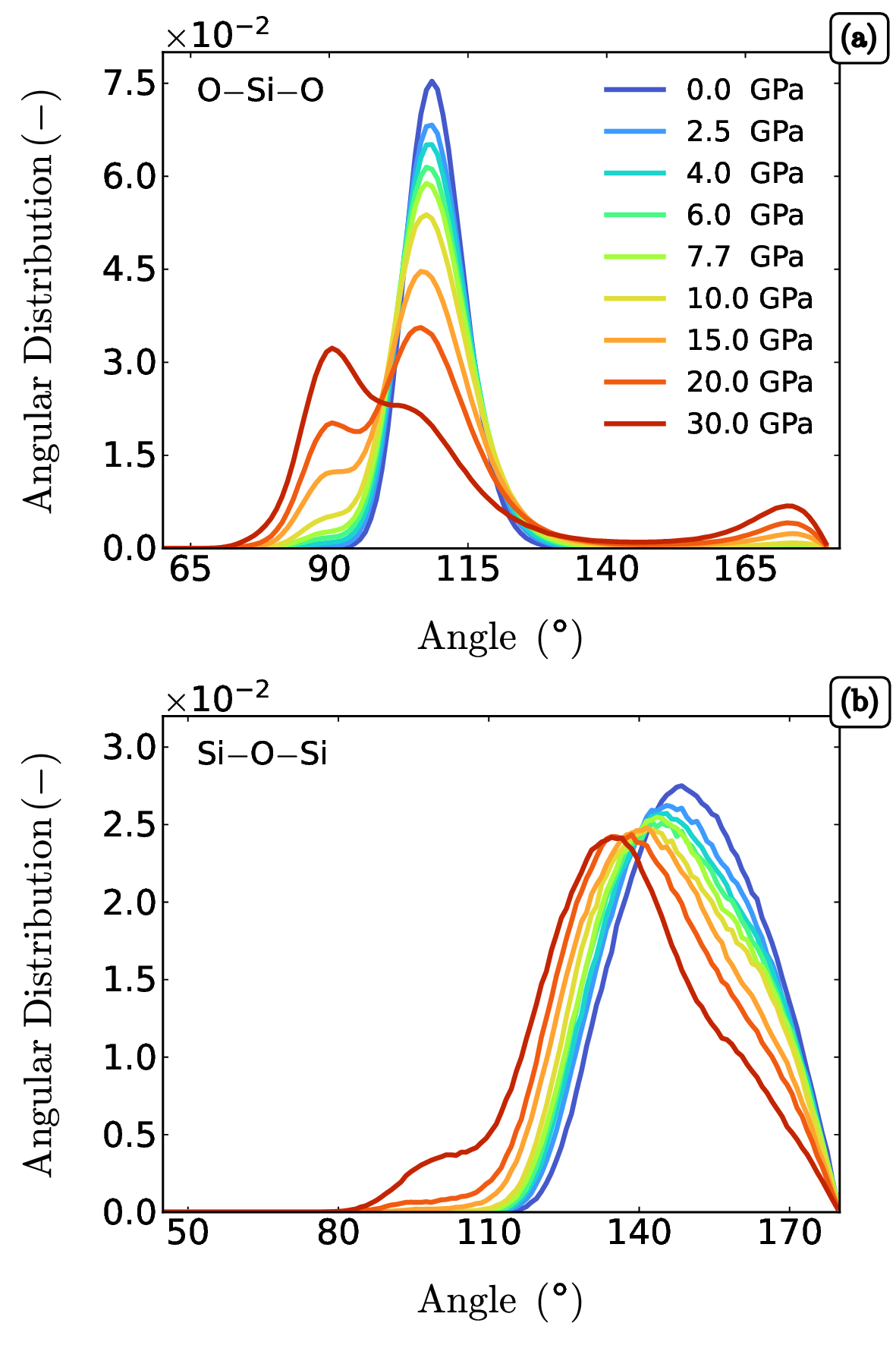}
\caption{(a) O--Si--O, (b) Si--O--Si Bond angle distributions as a function of pressure in LS$_2$ glasses at 300 K.}
\label{fig:adf}
\end{figure}

The separation distance between Si--O, Li--O, and O--O is given in Fig.~\ref{fig:BL}. These bond lengths were determined by fitting the first peaks of the radial probability density using the skewed-normal distribution (SND) presented in equation~\ref{eq1}.

\begin{equation}
\label{eq1}
f_{\text{SND}}^{(\mu,\sigma,\zeta)}(r) = \frac{e^{-(r-\mu)^2/2\sigma^2}}{\sqrt{2\pi\sigma^2}}\left[1+\erf\left(\zeta\frac{r-\mu}{\sqrt{2}\sigma}\right)\right],
\end{equation}
where $r$ is the distance between the elements, $\mu$, $\sigma$, and $\zeta$ denote the SND coefficients. This method has been established as an efficient approach for finding the bond lengths~\cite{Sukhomlinov17}. Figure.~\ref{fig:BL}(a) shows that the Si--O bond length decreases from 1.607 \text{\AA} to 1.587 \text{\AA} as the pressure increases, then it increases again to 1.589 \text{\AA}. The Li--O bond length is also plotted in the same figure alongside the contribution of BO and NBO to the total Li--O bond length (See Fig.~\ref{fig:BL}(a)). The Li--O bond length shows a decreasing trend with pressure, where the Li--O bond length displays a slight decrease from 1.931 \text{\AA} to 1.876 \text{\AA}, Li--NBO bond length decrease from 1.862 \text{\AA} to 1.802 \text{\AA}, and Li--BO bond length decrease from 2.128 \text{\AA} to 1.888 \text{\AA}. The O--O separation distance shown in Fig.~\ref{fig:BL}(b) displays the changes in O--O bond length with the pressure. We notice a decrease in the separation distance between the oxygen atoms, and the change goes from 2.614 \text{\AA} at 0 GPa to 2.517 \text{\AA} at 30 GPa. 
The CN of Si and Li atoms was found by counting the number of oxygen atoms within the first coordination shell, defined by a sphere with a radius equal to the distance where the first minimum of the g(r) is found. To ensure consistency, we maintained the same radius (2.1 \text{\AA} for Si--O, and 2.7 \text{\AA} for Li--O) for the CN calculation at high pressures as that determined at 0 GPa. 
The coordination numbers shows (Fig.~\ref{fig:CN}(a)) that the Si atoms have a 4-fold coordination number up to a pressure around 7.7 GPa and start increasing for pressures larger than it. At pressures lower than 7.7 GPa, most Si atoms are in a 4-fold coordination state (See Fig.~\ref{fig:xion}(a)). For pressures higher than 7.7 GPa, the amount of 5-fold and 6-fold coordinated Si atoms start to increase, with the amount of 5-fold coordinated Si atoms always being higher than the 6-fold Si atoms.
The behavior of Li is different, where initially, it has a 4-fold coordination in the glass prepared without external pressure and increases rapidly with pressure (goes from $~$4 to $~$6.3). 
The behavior of Li$^n$, with $n$ being the number of surrounding O atoms, is shown in Fig.~\ref{fig:xion}(b). At low pressure, 3-fold and 4-fold coordination states dominate and decrease with increasing pressure, while larger coordination states appear and dominate at high pressures. 
From the analysis of the neighboring atoms of the oxygen, two types of oxygen atoms were distinguished, bridging oxygen (BO), which is oxygen linked to two Si atoms, and non-bridging oxygen (NBO), which is oxygen linked to only one Si atom~\cite{Atila2019b}.
As illustrated in Fig.~\ref{fig:CN}(b), the dominant oxygen species in the LS$_2$ glass is BO, which is consistent with previous findings~\cite{Habasaki13, Atila2020c, Du2006}. We also observed an increase in the fraction of BO from 60.6\% up to 78.1\%, and a decrease in the fraction of NBO from 39.4\% to 18.4\% with increasing pressure.
To complement the short-range analysis depicted using the bond lengths and the coordination numbers, the bond angle distributions (BAD) will be used to gain more insight into how the pressure affected the LS$_2$ network topology. The BAD of O--Si--O and Si--O--Si are depicted in Fig.\ref{fig:adf}. In the LS$_2$ glass, the O--Si--O bond angle exhibits a peak at around 109$\degree$, which is close to the ideal tetrahedral angle of $\approx$ 109.4$\degree$. As the pressure increases, the peak of the O--Si--O bond angle distribution shifts towards lower angles around $\approx$106$\degree$. Also, we observed a shoulder at angles around 90$\degree$ that gradually became dominant in the LS$_2$ glasses for the highest cooling pressure. Furthermore, smaller peaks located closer to an angle of 170$\degree$ become more noticeable. The Si--O--Si BAD represents the angle between SiO$_4$ tetrahedra, and it has a peak at 148.5$\degree$ for the glass prepared with no external pressure and decreases to 134.5$\degree$ with a small shoulder appearing around 100$\degree$ in the glass cooled under a pressure of 30 GPa (see Fig~\ref{fig:adf}(b)).

\subsection{Medium-range structure}
\begin{figure}[h!]
\includegraphics[width=.98\columnwidth]{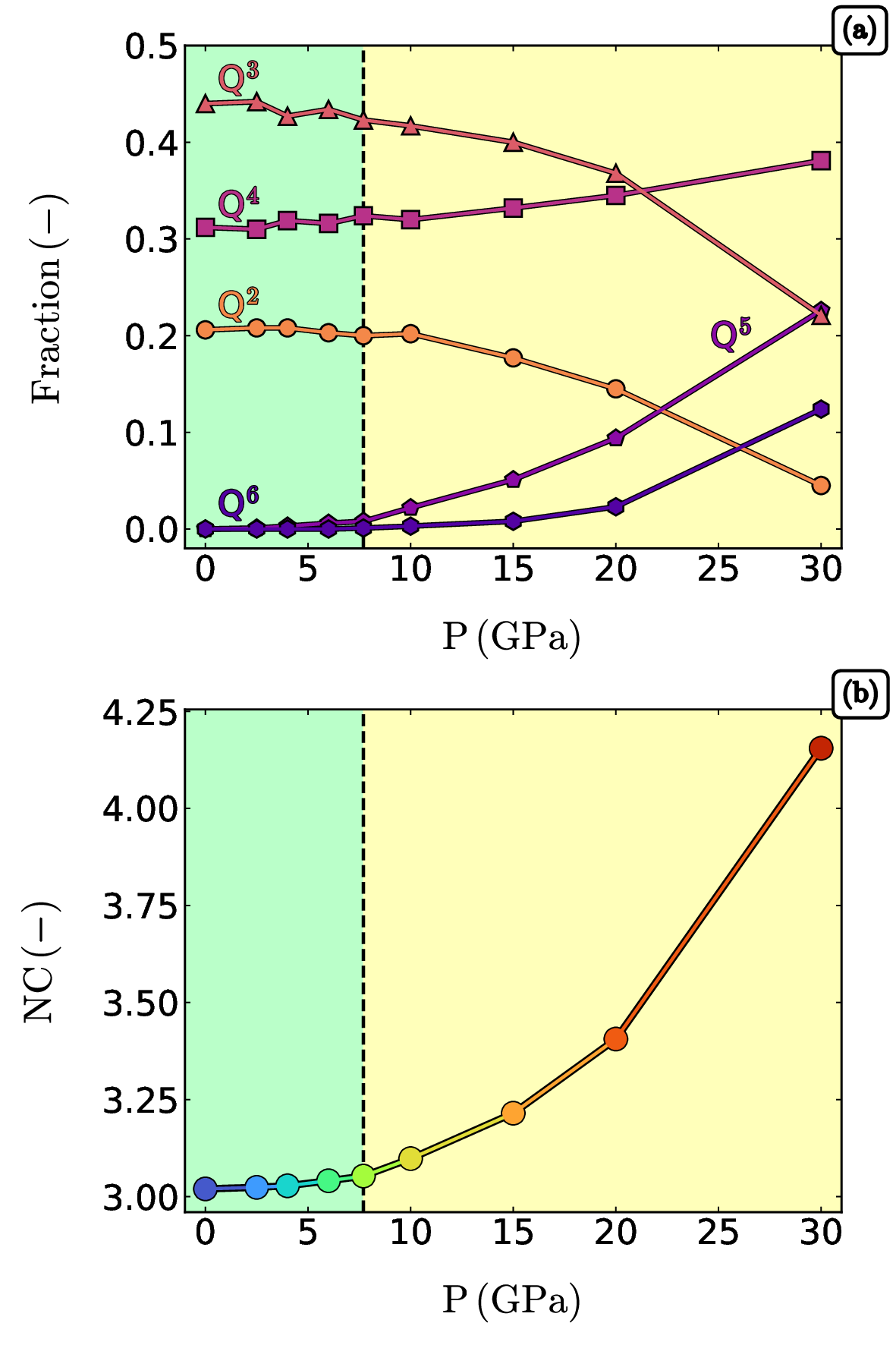}
\caption{(a) Q$^n$ distribution and (b) network connectivity as a function of pressure in LS$_2$ glasses at 300 K. Error bars are smaller than the symbol size.}
\label{fig:Conn}
\end{figure}

The medium-range structure of the silicate glasses can be seen as two (or more) connected tetrahedra and highlights the connectivity of the glass. The $Q^n$ distribution is a measure of how many BO or NBO are found in each tetrahedron, where the index $n$ refers to the number of bridging oxygen atoms present in a SiO$_4$ tetrahedron. Fig.\ref{fig:Conn} depicts the evolution of $Q^n$ percentages and the overall network connectivity of the LS$_2$ glasses as a function of the pressure. As pressure increases, the populations of $Q^2$ and $Q^3$ decrease while that of $Q^4$ increases. Additionally, $Q^5$ and $Q^6$ species significantly increase at higher pressures. Using the $Q^n$ distribution, we can calculate the network connectivity (NC) as the average number of bridging oxygen atoms per tetrahedron using equation \ref{eq2},
\begin{equation}
\centering
 \text{NC} = \sum_{i=1}^{i} nx_i
 \label{eq2}
\end{equation}
where $x_i$ represents the fraction of $Q^n$ species ($n$ $= 1$ to $6$). The NC tends to increase with pressure, ranging from 3.02 to 4.15, indicating that the glass structure is more polymerized at larger pressure. We note that from pressure below 7.7 GPa, almost no change was observed in the Q$^n$ population nor in the network connectivity.

\begin{figure}[h!]
\centering
\includegraphics[width=\columnwidth]{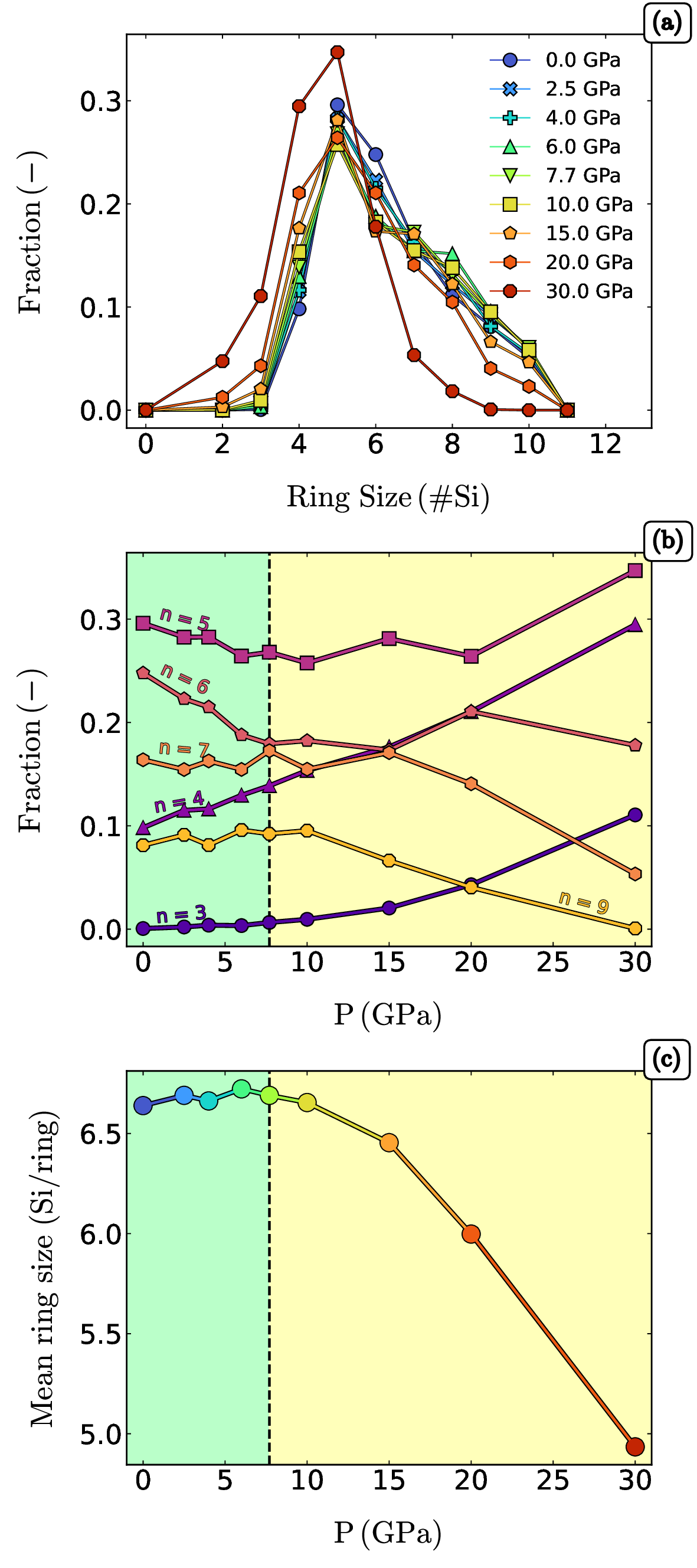}
\caption{Evolution of (a) Rings Size distribution, (b) Size of the rings, and (c) Mean ring size as a function of pressure in the LS$_2$ glasses at 300K. Error bars are smaller than the symbol size.}
\label{fig:RS}
\end{figure}

The next structural level of the medium-range order involves rings. The ring size distribution in the simulated LS$_2$ glasses was computed using the R.I.N.G.S code~\cite{RINGS} based on the Guttman criterion~\cite{GUTTMAN1990}, which was shown to give a realistic rings size distribution~\cite{Atila2019a, QiZhou2021}.
The size of a given ring is described by the number of Si atoms within it. The ring size distribution for the LS$_2$ prepared under no pressure (See Fig~\ref{fig:RS}(a)) is centered around five-membered rings, which is consistent with previous findings~\cite{Atila2020c}. The cooling under pressure significantly impacted the ring-size distribution. Fig.~\ref{fig:RS}(a) and (b) display a shift of the ring size distribution towards small ring sizes ($<$ five-membered) while the larger ones appear to diminish. However, the mean ring size shown in Fig.~\ref{fig:RS}(c) highlights no change in the ring topology for cooling pressure under 7.7 GPa, and then the mean ring size decreases markedly at higher pressures.

\subsection{Lithium clustering}
\begin{figure}[h!]
\centering
\includegraphics[width=\columnwidth]{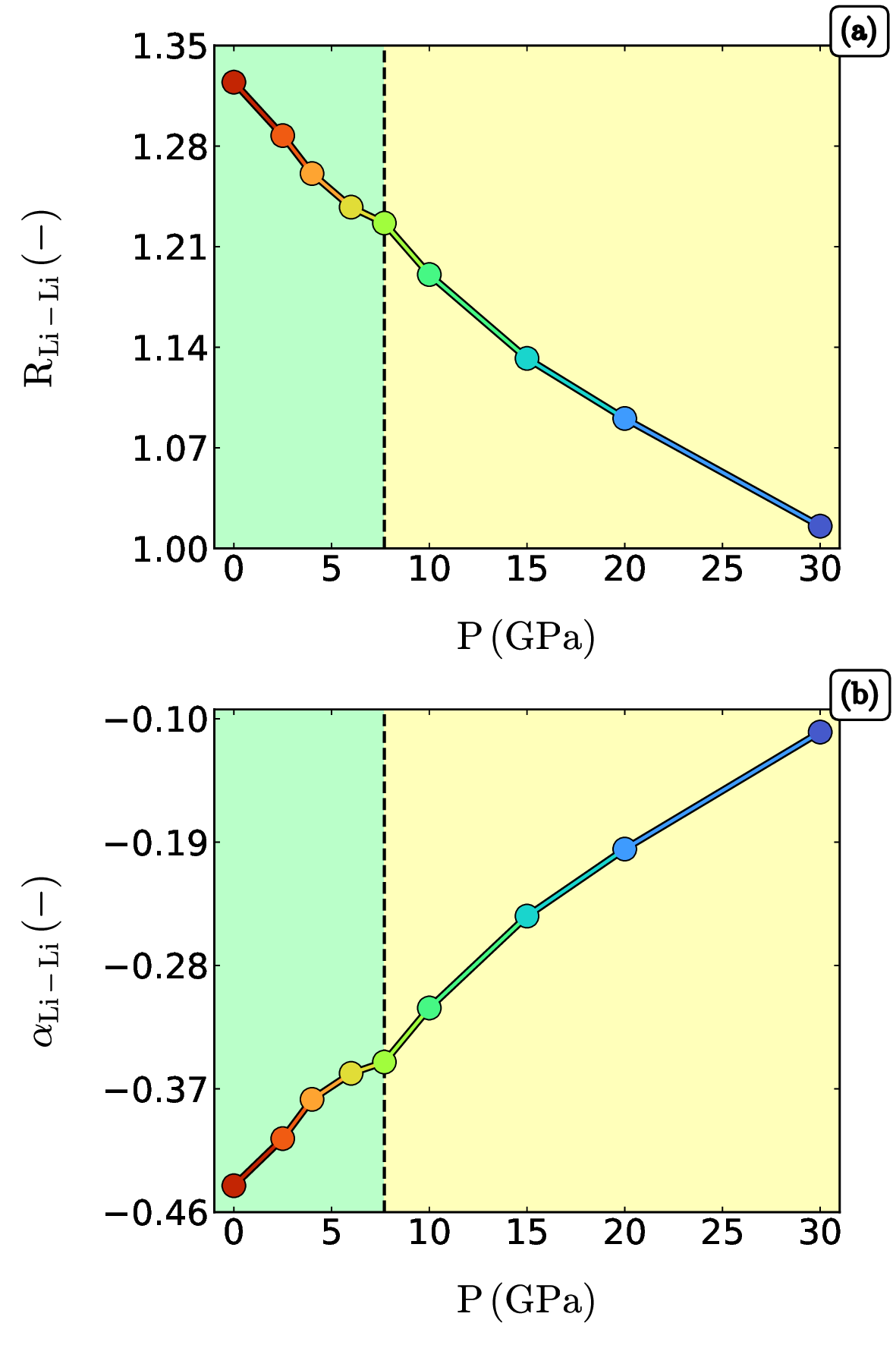}
\caption{(a) The clustering ratio of Li--Li pairs and (b) Chemical short-range order of Li as a function of the pressure at 300 K. Error bars are smaller than the symbol size}
\label{CR}
\end{figure}

\begin{figure*}[ht!]
\centering
\includegraphics[width=1\textwidth]{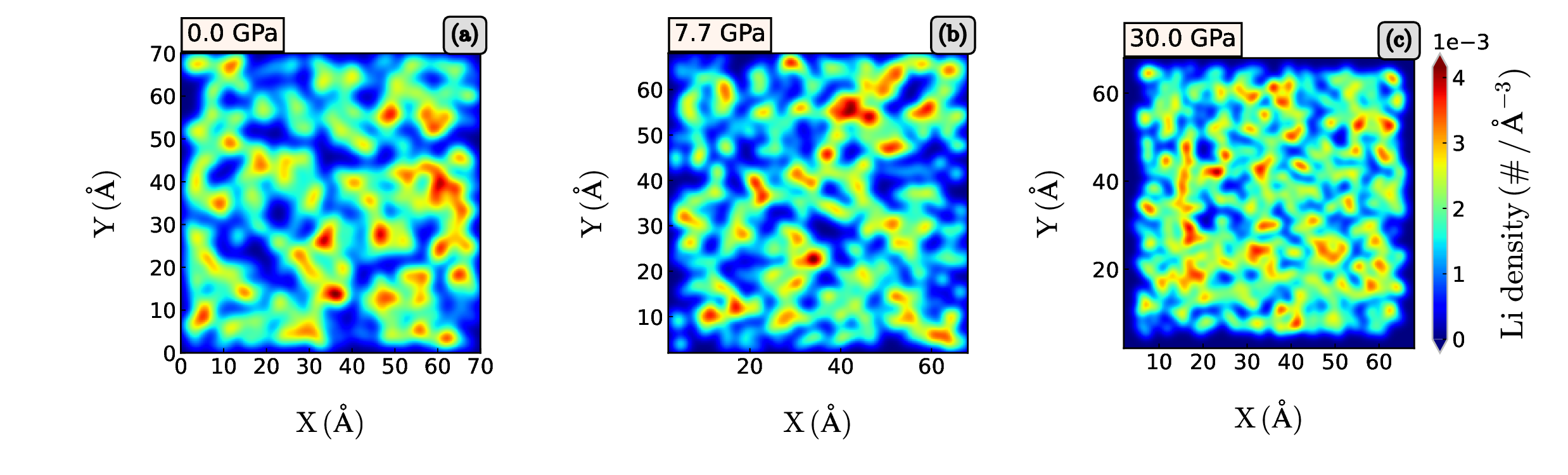}
\caption{The spatial distribution of lithium density at three selected pressures in a slice of the simulation box having a width of 8 \text{\AA} being twice the first minimum of the Li--Li g(r). Figure S1 in the supplementary materials shows the same property as in this figure but at different locations of the simulation boxes, highlighting the generalizability of the clustering behavior and its variation with pressure.}
\label{li_d}
\end{figure*}

In silicate glasses, modifiers tend to cluster and form channels, which causes some phase separation in the glass. We performed an analysis using the clustering ratio~\cite{Tilocca07}, calculated using eq.~\ref{eq3}:
\begin{equation}
R_{\text{Li}-\text{Li}}=\frac{CN_{\text{Li-Li}, \text{MD}}}{CN_{\text{Li-Li}, \text{hom }}}=\frac{CN_{\text{Li-Li}}}{\frac{4}{3} \pi r_c^3 \frac{N_{\text{Li}}}{V_{\text {box}}}},
\label{eq3} 
\end{equation} 
where R$_{\text{Li}-\text{Li}}$ is the clustering ratio, CN$_{\text{Li-Li}, \text{MD}}$ and CN$_{\text{Li-Li}, \text{hom}}$ is the Li--Li coordination number from MD simulations, and the Li--Li coordination number assuming a homogeneous distribution of Li atoms in the simulation box, respectively. $N_{Li}$ stands for the total number of Li atoms, $r_{c}$ the cutoff defined as the first minimum of the Li--Li g(r) (3.8 \AA), and $V_\text{{box}}$ the volume of the simulation box. Values of $R_{\text{Li-Li}}$ larger than 1 means that some clustering exists between the Li atoms. In contrast, $R$ = 1 indicates a statistical distribution of Li atoms throughout the sample. The change of $R_{\text{Li-Li}}$ with cooling pressure is shown in Fig.~\ref{CR}(a), where it decreases with increasing pressure. The clustering of Li atoms decreases with increasing pressure. This is also highlighted visually in Fig.~\ref{li_d}, where a density map of Li atoms in a slice of the simulation box is shown for selected pressures. Fig.~\ref{li_d} shows that in the LS$_2$ cooled with no external cooling pressure, clustering of Li atoms in the form of channel-like regions happens, while this clustering decreases with increasing cooling pressure.

In addition to the clustering ratio, we also computed the chemical of the short-range order (CSRO)~\cite{Cowley50} as defined in Eq.~\ref{eq4}: 
\begin{equation}
\alpha_{i-j} = 1 - \frac{\text{n}_{i-j}}{\text{N}_i \times \text{c}_j},
\label{eq4}
\end{equation}
where n$_{i-j}$ is the number of atoms of type j neighbors of a central atom of type $i$, N$_i$ is the total of neighbors for the atom of type $i$, c$_j$ is the concentration of type j. Generally, this parameter ranges between -1 and 1, with negative values indicating attraction and positive values highlighting repulsion. We can see clearly that the CSRO increases with increasing pressure, indicating the homogenization tendency of the Li ions in the LS$_2$ glasses in our study (see fig.\ref{CR}(b)).

\section{\label{Sec:Discussion}Discussion}
This study showcases the significant influence of pressure on lithium disilicate glass, as demonstrated through various analyses with changing pressure, which leads directly to a change in the glass density. 
Through the analysis of the structure factors, we were able to highlight that the topology of the short-range structure as depicted by the Si-O network (high $q$ values in Fig.~\ref{fig:xrsf} and Fig.~\ref{fig:sf}) is not affected by pressures lower than or equal to 7.7 GPa, where no changes were visible in the large $q$ regime. The change of the intensity of the FSDP and the correlation length with pressure show that we might have an amorphous--amorphous transition as an indication of polyamorphism, where we see an increase of the correlation length for cooling pressure lower than 10 GPa, and a decreasing correlation length for pressure higher than 10 GPa, which is counterintuitive if compared to metallic glasses under pressure~\cite{Atila2020b}, and inline with the work of Buchner \textit{et al.}~\cite{BUCHNER2014}. 

We observed high dependency between the change in the SiO$_x$ ($x =$ 4--6) units and the pressure (See Fig.~\ref{fig:xion}). In the glass cooled with no external pressure, it can be seen that the SiO$_4$ units were dominant; however, at pressures higher than 7.7 GPa, the SiO$_4$ are still dominant, but we find a significant increase in the SiO$_5$ and SiO$_6$ units in the glass which leads to a change in the glass topology. 
The formation of SiO$_5$ and SiO$_6$ units can also be observed through the O--Si--O and Si--O--Si angles, where the Si--O bond lengths experience a sudden increase between 20 and 30 GPa, while the CN of Si exhibits a more gradual increasing change. Notably, the O--Si--O angles exhibit a decrease to $\approx$ 90$\degree$, which is indicative of the characteristic angle of an octahedron. Another notable feature of this transition is an increasing peak at 170$\degree$, which becomes more prominent at higher pressure. This transition is a common feature in the silicate glasses~\cite{Ouldhnini2021, STEBBINS19, Kapoor17, Grujicic11}.

With increasing pressure, the network became more polymerized, which was indicated by the formation of more BO at the cost of NBO. This repolymerization was confirmed by the increase of the NC as the pressure increased, where Q$^n$ distributions are correlated with transformations of SiO$_x$ units as the pressure changes. 
The pressure effect on the ring-size distribution in the glass could be explained by the presence of topological variation in the structure, which was related to the type of modifier present in the glass~\cite{Atila2020c}. We notice that the mean ring size decreases with increasing pressure, indicating the dominance of small rings in the distributions; the dominance of the small rings ($<$ 6-membered) suggests the existence of small Li-rich regions in these glasses, which we can consider it as the first indication of Li homogenization in the glasses with increasing pressure~\cite{DU2004, Du2022}. 

The pressure-induced homogenization of the structure was shown using the clustering ratio and the density maps of the Li in the simulation box at different pressures (See Fig.~\ref{CR} and Fig.~\ref{li_d}). The increase of the homogenization of the Li distribution is mainly due to an increasing repulsion between the Li atoms when the pressure increases, which is confirmed by the analysis of the CSRO parameter~\cite{Cowley50}, where for the glass cooled under no external pressure, the Li atoms are more attracted towards each other, and this attraction decreases with increasing the pressure (See Fig.~\ref{CR}(b)). 
\section{\label{Sec:Conclusion}Conclusion}
In conclusion, by combining experiments and MD simulations, we showed that increasing pressure significantly impacts the structure of LS$_2$ glasses. polyamorphism was observed, with different amorphous structures with different correlation lengths. Moreover, observing the oxygen species accompanied by Q$^n$ distributions informs us that there is a high tendency for the repolmerization of the glasses as a function of the pressure. The ring size distribution suggests the existence of small Li-rich regions, which is an indicator of the homogeneity of the structure, which was shown visually through the density maps and quantitatively through the clustering ratio and CSRO. The presented results and insights into the pressure effect on the short-range order, structural unit connectivity, and homogeneity of LS$_2$ will help understand the glass-ceramics formation and control the crystallization.

\section*{Conflicts of interest}
There are no conflicts to declare.

\section*{Acknowledgements}
Y.B. thank the Centre National de la Recherche Scientifique et Technique (CNRST) for the financial support under the agreement number (N\degree: 3USMS2022). S.B., R.A.S., L.R., and A.S.P. would like to thank Conselho Nacional de Desenvolvimento Científico e Tecnológico (CNPq) and Fundação de Amparo à Pesquisa of the State of Rio Grande do Sul (FAPERGS) (project: 19/2551-0001978-5), Coordination for the Improvement of Higher Education Personnel for the financial support, and Laboratório Nacional de Luz Síncrotron (project 20170862).

\bibliography{apssamp}

\end{document}